\documentclass[11pt]{article}

\usepackage{amsfonts,amsmath,amssymb,graphics}

\title{\bf Universal trading under proportional transaction costs}

\newcommand{\R}{\mathbb{R}}

\newcommand{\typeA}{I}
\newcommand{\typeB}{II}

\newcommand{\ex}{\mathbf{E}}

\newcommand{\util}{\mathcal{U}}

\newcommand{\vZ}{\vec{Z}}

\newcommand{\notthis}[1]{}

\newcommand{\inv}{^{-1}}

\newcommand{\half}{\frac{1}{2}}

\newcommand{\money}{G}

\newcommand{\Time}{\textrm{time}}

\begin{document}
\author{Richard J. Martin\footnote{Dept.~of Mathematics, Imperial College London, SW7 2AZ, UK; Apollo Global Management International, 25 St George St, London W1S 1FS, UK}}
\maketitle

\begin{abstract}

The theory of optimal trading under proportional transaction costs has been considered from a variety of perspectives. In this paper, we show that all the results can be interpreted using a universal law, illustrating the results in trading algorithm design.

First published in RISK 27(8):54--59, 2014.

\end{abstract}


\section{Introduction}

In this paper we consider how to `optimally' deal with proportional
\footnote{The term `linear' costs often refers to the presence of fixed per-ticket cost and also a proportional part generated by a bid-offer independent of the trade size. As we are not considering a fixed part, we use the term `proportional', whereas in \cite{Martin11a} the term `linear' was used for the same thing.}
transaction costs when trading a single asset that follows an arbitrary diffusion process. Many of the superficial differences between the various strands of research are unimportant, and there is a universal law (eq.~\ref{eq:width})
which we formally publish here.
Although the literature on the subject is reasonably large, there is very little on applications in systematic trading algorithm design, so the purpose of this paper is to demonstrate it, with specific emphasis on the Sharpe ratio objective and variants of it, as these are the most often used in practice but have not been considered in the literature.

\begin{figure}[h!]
\begin{center}\begin{tabular}{c}
\scalebox{0.7}{\includegraphics*{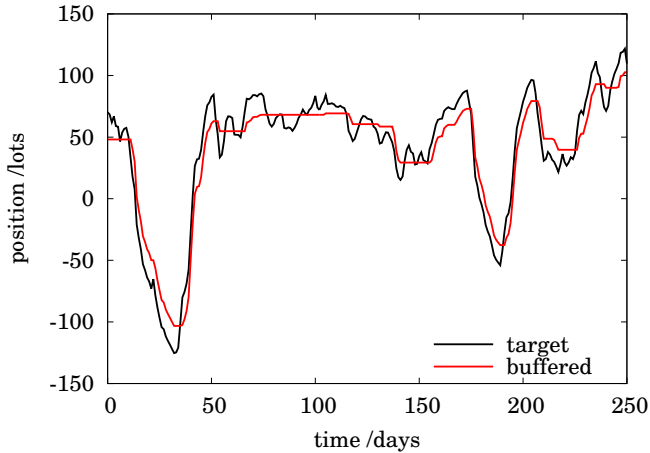}}
\end{tabular}\end{center}
\caption{\small Effect of buffering in example used later.}
\label{fig:tgtvsact}
\end{figure}

A systematic trading algorithm is a function that, given previous price history and/or the prices of other instruments, gives a `target position' ${\hat{\theta}}_t$ in the instrument to be traded. From the dynamics of these prices, which we assume diffusive, is inherited the dynamics of $\hat{\theta}_t$. The susceptibility of the strategy to transaction costs therefore depends principally on its volatility $\sigma_{\hat{\theta}_t}$.
As the target position changes by $O(dt^{1/2})$ over a time period $dt$, money will be lost at an infinite rate. It is well known (see initially \cite{Magill76}) that the optimal strategy is to draw a `buffer' around the target position, defining a no-trade (NT) zone in which the position is held unchanged and on each side a discrete-trade (DT) zone in which one trades immediately to the edge of the NT zone:
see Figure~\ref{fig:tgtvsact}. This prevents the strategy constantly trading backwards and forwards: typically the action is to restrict trading to a succession of small trades in one direction only, subsequently reversing. The question we are to address is, what is the optimal buffer width? Too narrow, and one loses too much in costs by overtrading; too wide, and the so-called `displacement loss', as a result of having a non-optimal position on, is excessive (in option delta-hedging problems the delta would be too far from zero).

\subsection*{Optimality}

We should first define what optimality means. Essentially, there are two types of problem, which we call Type \typeA/\typeB:
\begin{itemize}
\item[\typeA] Traditional expected utility maximisation of terminal wealth with or without consumption, such as the so-called `Merton problem' of rebalancing a portfolio of a stock and riskfree bond, or the delta-hedging of options;
\item[\typeB] Local utility maximisation \cite{Kallsen99}, as applied in systematic trading algorithms \cite{Martin11a,Martin12a}. This means that incremental rather than terminal P\&L variation is being penalised.
\end{itemize}
There are two different methods of solution: maximise the expected utility with respect to the buffer width, and understand how the utility depends on $\varepsilon$ at leading order; or obtain the exact boundary and then consider how that depends on $\varepsilon$.

\subsection*{The optimal buffering formula, and dependencies}

In a working-paper \cite{Martin12a}, dealing with Type \typeB, we state that the optimal half-width, in the limit of small costs, depends on only four quantities: the volatility of the target position $\hat{\theta}_t$ and the (absolute, \$) volatility of the  underlying asset $X$, the ratio of these being denoted $\hat{\Gamma}_0$; the proportionality coefficient of cost, $\varepsilon$; and a parameter pertaining to the degree of absolute risk aversion, denoted $G$ and having monetary units. The formula for the half-width $\delta\theta$ is
\begin{equation}
 \delta\theta  \sim \left(\frac{3\varepsilon \money \hat{\Gamma}_0^2}{2}\right)^{1/3},
\qquad \hat{\Gamma}_0^2 
= \frac{\sigma^2_{\hat{\theta}}}{\sigma^2_X} .
\label{eq:width}
\end{equation}
The same result is obtained in a working-paper of Kallsen \& Muhle-Karbe \cite{Kallsen13} (an excellent discussion of the subject and its literature), using Type \typeA.
The optimal buffer is also displaced from the no-cost position, reflecting the fact that if one needs to buy more of the asset to reach the target position, but yet the target position is predicted to decrease, then one should buy less now. This result is not universal and as the displacement is typically small we ignore it in this paper.

\notthis{
\begin{equation}
\mathrm{d}\theta 
\sim
\frac{\mu_{\hat{\theta}}}{\sigma_X^2} \left(\frac{2\varepsilon^2G^2}{3\hat{\Gamma}_0^2} \right)^{1/3} ,
\label{eq:displ}
\end{equation}
which is smaller, being only $O(\varepsilon^{2/3})$, and we ignore it. Loosely, it anticipates the change in position (if one has no position on, and the target position is positive, then one should not buy as much if one knows that it is likely to reduce). When it is not ignorable, the transaction costs and the buffering are so wide as to make the trading strategy ineffective anyway. It remains conjectural for problems of Type \typeA\mbox{} because $\mathrm{d}\theta$ has no influence at leading order, so it is hard to infer.
} 

The application of this results is simple because the parameters are easily attributable to obvious characteristics of the trading strategy; we give a couple of calculations in a moment.

The ratio $\hat{\Gamma}_0^2$, which in option delta-hedging problems is just the square of the option's gamma, is understood as an attempted trading speed: the higher it is, the wider the buffer. The cube-root dependence on $\varepsilon$ is important. For single-period problems it is immediate that the law is actually a proportional one; but this ignores the fact that when many trades are done, at potentially very small time intervals, the frequency of trading has to be taken into account, with more trades causing more loss. By wrongly using a proportional law, one ends up overtrading cheap markets, incurring excessive losses through transaction cost,  and  undertrading expensive ones, causing excessive displacement loss. 

Only $G$, which has dollar units and is application-dependent, requires further explanation. In Type \typeA\mbox{} problems, $G$ is the scaling constant in the utility function and specifies what variability of terminal wealth is acceptable. In Type \typeB\mbox{} problems, the target position is proportional to the instantaneous rate of return of the asset and inversely proportional to the square of its volatility: it is in fact $G$ the desired dollar P\&L volatility per unit Sharpe of the strategy \cite[Eq.6]{Martin12a}, the idea being that one allocates a `risk budget' proportional to one's financial means and to the expected Sharpe ratio, and the effect of changing $G$ is simply to rescale the position: $\hat{\theta} = G \mu_X/\sigma_X^2$. 

As an example, take the trading of the front Treasury note futures contract TY1, about which we assume the following: spot price 127, bid-offer 0.02, 1pt = \$1000, Black-Scholes ATM vol $5\%$.  
So then $\sigma_X=$ \$400 (= typical \$ daily move of one contract), and $\varepsilon=$ \$10 (=half the bid-offer of one contract, in \$). Now let the strategy be geared at $G=$ \$1,000,000 and suppose that its daily position changes by typically 35 contracts. Then 
\[
\delta \theta = \left( \frac{3}{2}\times  \$10 \times \$10^6 \times \frac{35^2}{(\$400)^2} \right)^{1/3} = 49 \;\mbox{contracts}.
\]
Notice the need to convert everything to consistent units: $X$ is written in dollars, $\theta$ is a number of contracts, time in business days (necessitating conversion from annualised to daily vol). 

As a second example, take the trading of a CDS index contract, spread 80bp, vol 35\%, 5Y duration, bid-offer 0.75bp. One lot ($\theta=1$) means a notional of \$1,000,000. Then $\sigma_X$ = \$875 (daily) and $\varepsilon=$ \$190. Let the strategy be geared at $G=$ \$500,000 and suppose that its daily position change is typically \$4,000,000. Then
\[
\delta \theta = \left( \frac{3}{2}\times  \$190 \times \$500\cdot10^3 \times \frac{4^2}{(\$875)^2} \right)^{1/3} = 14 \;\mbox{(\$M notional)}.
\]


\subsection*{Universality}

The most important thing is what gives rise to instantaneous variation of the target position, which is why $\hat{\Gamma}_0$ enters: that is what directly gives rise to transaction losses. Many ingredients in the setup do not signify, because they do not convey information beyond that of $\hat{\Gamma}_0$, for example:
\begin{itemize}
\item
whether the objective function is utility-based or Sharpe-based (we explain this in more detail presently);
\item
the utility function, or the risk measure;
\item
whether the objective function pertains to a finite or infinite horizon;
\item
whether the asset is `cash-like' (like a stock following a geometric Brownian motion), or `synthetic' (as in for example the PV of a swap, hence written down as an arithmetic Brownian motion with possibly non-constant volatility);
\item
the origin of the problem, e.g.\ option hedging, systematic trading;
\item
the number of factors used to describe the asset dynamics.
\end{itemize}
Thus (\ref{eq:width}) is universal, with the proviso that the symbols may need reinterpretation from case to case: for example in equity trading, deciding whether $\theta$ means the \$ notional of a stock, or whether it means the number of shares. This is just a question of ensuring consistency.

To put these into context, and to justify universality, here is a brief overview of published and unpublished results that links everything up:

\subsubsection*{Option delta-hedging: \cite{Whalley97,Zakamouline06} (Type \typeA)}

Generally the objective function is utility of terminal wealth and the horizon is the option expiry. As the terminal P\&L can be negative, CARA i.e.\ negative-exponential utility is used. Their formula \cite[Eq.12]{Zakamouline06} is as stated in (\ref{eq:width}), if we substitute $\varepsilon=\varepsilon' S$, $\sigma^2_{\hat{\theta}}/\sigma^2_X=\sigma^2S^2\Gamma^2/\sigma^2S^2$ (thus $\hat{\Gamma}_0^2=\Gamma^2$, explaining our choice of notation), with $S$ the stock price, $\sigma$ the Black-Scholes volatility, $\Gamma$ the option gamma, and $\varepsilon'$ the transaction cost as a fraction of the stock price. We also write $\money$ in terms of their risk aversion coefficient $\gamma$ via $\money=e^{-r(T-t)}/\gamma $, the discounting coming from the fact that $G$ pertains to time $t$ (today) and $1/\gamma$ to time $T$ (expiry).  The resulting formula $ \delta\theta$ is the width of the NT zone in delta units, as opposed to a \$ notional of stock.  

\subsubsection*{Systematic trading with `zero-factor' model:  \cite{Martin11a} (Type \typeB)}

(See also \cite{Rej15}.) We formulate systematic trading models by expressing the dynamics of the traded asset as functions of `factors', $\vZ$:
\[
dX_t = \mu_X(\vZ_t)\, dt + \sigma_X(\vZ_t)\,dW_{X,t};
\]
these factors may be exogenous (e.g.\ coming from analyst views, or market prices of other instruments) or endogenous (e.g.\ momentum). By zero-factor we mean a one-factor model in which the factor $Z_{1,t}$ is actually the asset $X_t$ being traded: $Z_1\equiv X$. (A general one-factor model therefore has \emph{two} moving parts: the factor and the traded asset.) The objective function is not utility of terminal wealth, but utility of \emph{changes in} wealth---the so-called local utility functions, in which the objective is $\util(\theta_t \,dX_t)$ summed, with $\util$ a smooth concave function. Without loss of generality we can impose $\util(0)=0$, $\util'(0)=1$, $\util''(0)=-1/\money$, and for diffusive dynamics it boils down to a quadratic objective. The objective (value function) is
\begin{eqnarray}
V_t &=&  \ex_t \left[ \int_{s=t}^\infty  e^{-r(s-t)} \util(\theta_s  \, dX_s) \right]   \nonumber \\
&=& \ex_t \left[ \int_{s=t}^\infty  e^{-r(s-t)} \bigg(\theta_s \mu_{X_s} - \frac{\theta_s^2 \sigma^2_{X_s}}{2\money} \bigg) \, ds\right] ,
\label{eq:objfn}
\end{eqnarray}
so the target position is $\hat{\theta}_t = \mu_{X_t}G/\sigma_{X_t}^2$. The infinite horizon causes the value function to obey an ordinary differential equation in $X$ rather than a parabolic PDE in $(t,X)$.  Although this particular model has limited application, it has the remarkable consequence of a reasonably explicit solution for the NT boundary, obtained in \cite{Martin11a}, though the equations are unwieldy; also, it forms the basis for the discussion of multifactor models which are very general. Note that \cite{Martin11a} gives a non-heuristic derivation of the cube-root law, because it drops out of a Taylor series expansion of the solution\footnote{It balances the two leading order terms, which turn out to be $O(\delta\theta)^3$ and $O(\varepsilon)$.}: thus one does not need to `know in advance' that expansion in powers of $\varepsilon^{1/3}$ was necessary. 
The simplest case is where $X$ follows Ornstein-Uhlenbeck dynamics $dX_t=-bX_t\,dt+\sigma\,dW_t$:
\[
\hat{\theta} = -bXG/\sigma^2, \qquad  \delta\theta  \sim (3\varepsilon b^2 / 2 \sigma^4 )^{1/3} \money .
\]
Bouchaud and co-workers corroborate this by different techniques \cite{Lataillade12}.

\subsubsection*{Systematic trading with multifactor model:  \cite{Martin12a} (Type \typeB)}

When many factors cause the target position to vary, the position is still $\hat{\theta}_t = \mu_X(\vZ_t)G/\sigma_X(\vZ_t)^2$ which is a function of $\vZ\in\R^m$, but despite the higher dimensionality, eq.~(\ref{eq:width}) still holds. This is because of `locality' again. In the $(m-1)$-dimensional space perpendicular to the gradient vector $\nabla \hat{\theta}(\vZ)$, the target position does not change for small changes in $\vZ$, and so no transaction costs occur: one only needs to study the direction of greatest variation in $\hat\theta$, which is $\nabla \hat{\theta}(\vZ)$, reducing everything to a one-dimensional problem.

\subsubsection*{CRRA utility (Type \typeA)}

Constant relative risk aversion is applicable to what might be described as cash strategies, such as the rather academic `Merton problem' of rebalancing a portfolio of stock and bond \cite{Davis90,Shreve94}. It only makes sense when the portfolio value is always $\ge 0$ (as the utility function is singular at 0). As pointed out by Kallsen \& Muhle-Karbe \cite{Kallsen13}, one replaces the gearing $G$ with $W/\gamma$, where $W>0$ is the current wealth and $\gamma$ the coefficient of constant relative risk aversion.
That $G$ is no longer constant does not matter: this variability is not a local effect. It simply says that in the future, when I have made or lost money, I will want higher or lower gearing, but that is irrelevant to the question of how to mitigate the transaction costs associated with trading occurring \emph{now}.

\subsubsection*{Utility of terminal wealth vs Local utility (Type \typeA/\typeB)}

Utility of terminal wealth (with or without a consumption term) just about makes sense in investment problems, but not in systematic trading, for two reasons. First, there is no well-defined time horizon, and investors are typically worried about short-term variation in P\&L, particularly as that gives rise to drawdowns. Secondly, although such strategies can be tested with simulated data (as can the option-hedging problems) by running many simulations, they cannot legitimately be backtested on real data as there is only one trajectory from which to form the expected utility. On the other hand, with a couple of decades' data, one can form an opinion about the Sharpe ratio if risk is taken as \emph{variation} in P\&L, or again the so-called local utility functions (q.v.): these are natural in the trading world, as they relate easily to a `daily VaR limit'. 
Nonetheless, the buffering law is the same, and again this is an argument about `locality': whatever the objective function, the important thing is the variability of the target position.

\subsubsection*{Sharpe ratio vs local utility; different risk measures (Type \typeB)}

It is easily established (see e.g.~\cite{Martin12a}) that the classic Markowitz optimisation problem of maximising expected return subject to an upper bound on the risk, to be interpreted as quadratic variation of P\&L, is equivalent to (\ref{eq:objfn}) above; the proof is a simple exercise in Lagrange multipliers, with $G$ in (\ref{eq:objfn}) being the reciprocal of the Lagrange multiplier. Therefore, Sharpe ratio optimisation, in which we maximise mean $\div$ square root of quadratic variation, is equivalent to the local quadratic utility formulation of (\ref{eq:objfn}).

If the market is diffusive, and our estimate of volatility $\hat{\sigma}_{X_t}$ is correct, there is no difference between using standard deviation of daily returns (to compute the local utility) or some other risk measure such as VaR or shortfall. In practice, these assumptions are dubious, but in the absence of any theoretical results for non-diffusive markets, we may as well attempt to use (\ref{eq:width}). The numerical results that we presently show suggest that this is justifiable. This is unsurprising, because losses from transaction costs largely arise from the diffusive component of the market returns, not from occasional jumps.


\section{Numerical demonstration}


We demonstrate eq.~(\ref{eq:width}), choosing Sharpe ratio as a performance measure. In principle we must test against all other buffering schemes; and another difficulty is that the buffer width is generally time-varying, so we cannot simply plot Sharpe ratio vs buffer width (there is no unique buffer width to plot on the horizontal axis). What we can do, though, is multiply $\delta\theta$ in eq.~(\ref{eq:width}) by some fixed amount $\lambda$, and plot the \emph{time-average} of the buffer width on the horizontal axis, and  on the vertical axis the Sharpe ratio. Repeating for different values of $\lambda$ causes a curve to be described, and we highlight the point corresponding to $\lambda=1$. Finally, we repeat for different transaction cost parameters to give a family of curves. 

Consider what this curve should look like, as a function of $\lambda$. If the buffer width is too small ($\lambda\to0$) then too much value will be lost, and in continuous time it would drop to $-\infty$: the drop will be severe if $\varepsilon$ is high. If the buffer is too wide, the displacement loss takes over and the performance should drop. Indeed, in the limit  $\lambda\to\infty$ the NT zone will become so large that no trading takes place at all, and then the Sharpe ratio will become undefined. At some intermediate point there should be a maximum and ideally the result for $\lambda=1$---which we mark in the Figures---will be exactly there, indicating that no improvement can be made by scaling (\ref{eq:width}) up or down by a fixed amount; though it does not rule out the possibility that the buffer is suboptimal by virtue of being at some times too wide and at other times too narrow. However, if the costs are high enough, the value function will always be negative and there will be no hump: then the strategy is worthless, irrespective of how well `optimised' the buffer is.

For these models the gearing plays no useful role because it simply scales the position, the buffer width, the expected P\&L, and the risk, all in direct proportion, so it has no effect on the Sharpe ratio. We therefore set $G$ to \$1M throughout.
We are going to use VaR and shortfall (ESF) as well as standard deviation in the Sharpe ratio calculation. We fix the tail probability as $p=0.01$, and for convenience we divide the VaR by $\Phi\inv(1-p)$ and the ESF by $\phi(\Phi\inv(p))/p$ so that\footnote{$\Phi$, $\phi$ denote as usual the standard Normal cdf and pdf.} for Normal distributions of zero mean all these measures are identical. Fat-tailed return distributions thereby produce lower VaR-Sharpe and ESF-Sharpe---as is seen here, though the effect is slight.

Note incidentally that the alternative method of finding the optimal buffer width is dynamic programming. This is impractical unless the number of factors is small: in real trading algorithms one may well have ten or so factors, requiring an optimisation in ten dimensions.

\subsection*{Examples using synthesised data}

Synthesised models allow arbitrarily much data to be generated, with all parameters known. We consider the one-factor linear model,
\begin{eqnarray}
dX_t &=& \beta\sigma_X  Z_{1,t} \, dt + \sigma_X \, dW_{0,t} \label{eq:modlin} \\
dZ_{1,t} &=& -\kappa Z_{1,t} \, dt + \sqrt{2 \kappa} \, dW_{1,t} \nonumber 
\end{eqnarray}
with Type \typeB\mbox{} optimisation. The factor $Z_1$ follows a standardised OU process and is understood as a sort of bull-or-bear indicator: when positive, $X$ drifts upwards, and when negative, downwards. 
It is immediate that
\[
\hat{\theta}_t = \frac{\beta Z_{1,t}\money}{\sigma_X} , \qquad
\hat{\Gamma}_0^2 = \frac{2\beta^2 \kappa G^2}{\sigma_X^4},
\]
and that the Sharpe ratio for $T$-period trading returns is $|\beta| T^{1/2}$, with which simulation should agree (and does).
Thus
\[
\delta\theta \sim \left(\frac{3\varepsilon\kappa}{\sigma_X|\beta|}\right)^{1/3} \frac{\money |\beta|}{\sigma_X} 
= 
 \left(\frac{3\hat{\varepsilon}\kappa}{|\beta|}\right)^{1/3} \big( \overline{\theta^2} \big)^{1/2} 
\]
where $\hat{\varepsilon}=\varepsilon/\sigma_X$ is the cost per unit volatility of the tradable and $(\overline{\theta^2}) ^{1/2} = \money\beta/\sigma_X$ is the root mean square position (not the same as $\sigma_{\hat{\theta}}$ which pertains to \emph{changes} in position).  
Notice that in this simple case $\hat{\Gamma}_0^2$ and hence $\delta\theta$ are constant---though they are not if the coupling is nonlinear, i.e.\ in the drift of $dX_t$ we replace $Z_1$ by a function $\psi(Z_1)$---and that correlation between $dW_{X,t}$ and $dW_{Z_1,t}$ does not play a part. Notice also that the buffer width and position are both inversely proportional to $\sigma$, provided one fixes $\hat{\varepsilon}$. (If the volatility of the underlying increases with $\varepsilon$ fixed, then the asset has actually become cheaper to trade and the buffer width drops as a fraction of the typical position.) Thus the only factors that link the buffer width to the r.m.s.\ target position are $\hat{\varepsilon}^{1/3}$ and an extra quantity $\kappa/|\beta|$ that has dimensions $\Time^{-1/2}$; this is necessary for dimensional agreement (because $\hat{\varepsilon}$ has dimensions $\Time^{1/2}$) and can be thought of as the trading speed, because the higher $\kappa$ is the more rapidly the factor is changing direction. Finally, if $\beta\to0$ then the buffer width becomes large as a fraction of the r.m.s.\ position (not in absolute terms because the r.m.s.\ position reduces too): the explanation for this is that the asset price has in effect become less predictable, or that the trading signal is of lower quality: as expected, therefore, the NT zone becomes relatively wide and cuts down the amount of trading.

\begin{figure}[h!]
\begin{center}\begin{tabular}{c}
\scalebox{0.65}{\includegraphics*{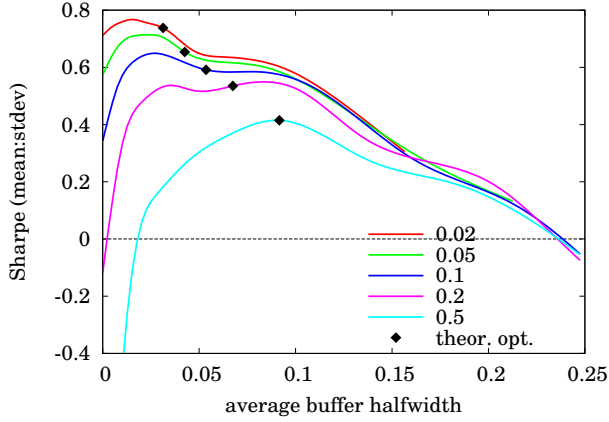}}
\end{tabular}\end{center}
\caption{\small Performance vs buffer size for synthesised model. Cost multipliers ($\varepsilon$) are indicated on graph. Buffer size is as notional in \$M. Theoretical optimum marked in each case.}
\label{fig:perf_lin}
\end{figure}

Figure~\ref{fig:perf_lin}(a) shows results with\footnote{Dimensions of $\kappa$, $\beta$, $\sigma$ are respectively $\Time^{-1}$, $\Time^{-1/2}$, $\$/\Time^{1/2}$, where units of $\Time$ need to be consistent throughout: we are having them as business days. 10,000 data points were used.} $\kappa=0.02$, $\beta=0.04$, $\sigma=0.5$, $\rho_{01}=0$, for transaction cost $\varepsilon=$ 0.02, 0.05, 0.1, 0.2, 0.5. 
 The appearance of the graphs is as expected and the postulated rule (\ref{eq:width}) appears to be optimal. For low costs the impact of getting the buffer wrong is quite small, but for high costs it is much bigger: being out by a factor of 2 makes a huge difference in performance. As returns are Normally distributed here, the results for VaR and ESF are identical and hence are omitted.

Extensions to nonlinear coupling, multiple factors and stochastic volatility are given in \cite{Martin12a}: the results are all pretty much the same, suggesting that the above model contains all the important ingredients.

\subsection*{Examples using real data}

The model of the previous section is most easily interpreted as an exogenous factor driving the tradable asset, and the model parameterisation can be determined by observation, subject of course to estimation error. Here the construction is different and there is the additional problem that once does not know the underlying model: but as it turns out, that does not matter.

In momentum models, the factors are implicit, and estimated using moving averages of the asset being traded. 
Using integration by parts, a moving average of prices can be reexpressed as a weighted sum of returns,
\[
Z_{t} = \int_{\tau=-\infty}^{t} K(t-\tau) \, dX_{\tau}/\sigma_{X_\tau}.
\]
For the commonly-used exponentially-weighted moving average crossover,
\[
K(\tau)=\frac{\sqrt{2(T_\textrm{s}+T_\textrm{f})}}{|T_\textrm{s}-T_\textrm{f}|} \big(e^{-\tau/T_\textrm{s}} -e^{-\tau/T_\textrm{f}}\big)
\]
\notthis{
The commonly used exponentially-weighted moving average crossover is defined as\footnote{In \cite{Martin12c} gives a reformulation using the returns $dX_\tau$.}
\[
Z_t = \frac{T_\textrm{f}\inv \int_{-\infty}^{t} e^{-(t-\tau)/T_\textrm{f}} X_\tau \, d\tau - T_\textrm{s}\inv \int_{-\infty}^{t} e^{-(t-\tau)/T_\textrm{s}} X_\tau \, d\tau}{\hat{\sigma}_{X_t} \cdot |T_\textrm{s}-T_\textrm{f}| \big/ \sqrt{2(T_\textrm{s}+T_\textrm{f})}}
\]
} 
where $T_\textrm{f}$ and $T_\textrm{s}$ are the `periods', so for example a 5:10 day moving average has $T_\textrm{f}=5$, $T_\textrm{s}=10$. The prefactor normalises $\overline{Z^2}=1$.
It is also advantageous \cite{Martin12c} to transform the normalised signal using a `response function' $\psi$ rather than simply having a position proportional to $Z$.
By combining filters of different speeds, one obtains a prediction of $dX_t$, and thence the target position:
\[
\hat{\theta}_t = \frac{\beta_1\psi(Z_{1,t}) + \cdots + \beta_m\psi(Z_{m,t})}{\sigma_{X_t}}.
\]
The factor weights $(\beta_j)$ are fitted by regression or by optimising the backtested performance, or can be set manually. For these purposes we use 2:4, 4:8, 8:16, 16:32 day crossovers and the response function is $\psi(z)=ze^{-z^2/2}$. Now, it is difficult to get $\hat{\Gamma}_0^2$ theoretically, but very simple to estimate it empirically from the observed quadratic variation of $\hat\theta$ and $X$ in the natural way:
\[
{(\hat{\Gamma}_0^2)}_t \approx \frac{
\sum_{n=0}^\infty \alpha^n (\hat{\theta}_{t-n\, \delta t}-\hat{\theta}_{t-(n+1)\, \delta t})^2
}{
\sum_{n=0}^\infty \alpha^n (X_{t-n\, \delta t}-X_{t-(n+1)\, \delta t})^2
},
\]
with $\alpha$ being the `forgetting-factor' (we used an effective `period' of 32 days in simulation, so $\alpha=1-\frac{1}{32}$).


In the example we have considered here the time series of the traded asset $X_t$ is given by stitching together the time series of the individual futures contracts\footnote{This can be done automatically in Bloomberg ({\tt GFUT <Go>}). We adjust fixed-income contracts by difference and everything else by ratio.}. The time series is assumed to exhibit trending to some extent, which should result in P\&L generation from a momentum strategy. Two contracts are used: TY1 (front US Treasury note futures) and RR1 (rough rice). 
Positions and buffer sizes are expressed in \$M notional; for TY, costs are as a fraction of par\footnote{e.g.\ if market is 128-03/128-03+ then $\varepsilon=\half\times\frac{1}{64}/100\approx0.0001$. US Treasuries are quoted in $\frac{1}{32}$s with + denoting $\frac{1}{64}$.} and for RR they are a proportion of the current futures price\footnote{e.g.\ if market is 20.16/20.18 then $\varepsilon=\half\times0.02/20.17\approx0.0005$.}.
Again we fix $G=$\$1M.
The position and the buffer size are expressed as notional amounts in \$M: to express them as a number of contracts, just divide by the contract size\footnote{e.g.\ TY: this is \$100,000. Thus $\theta$ = \$32M means 320 contracts.} in \$M.

\begin{figure}[h!]
\hspace{-20mm}
\begin{tabular}{ll}
(a) \scalebox{0.55}{\includegraphics*{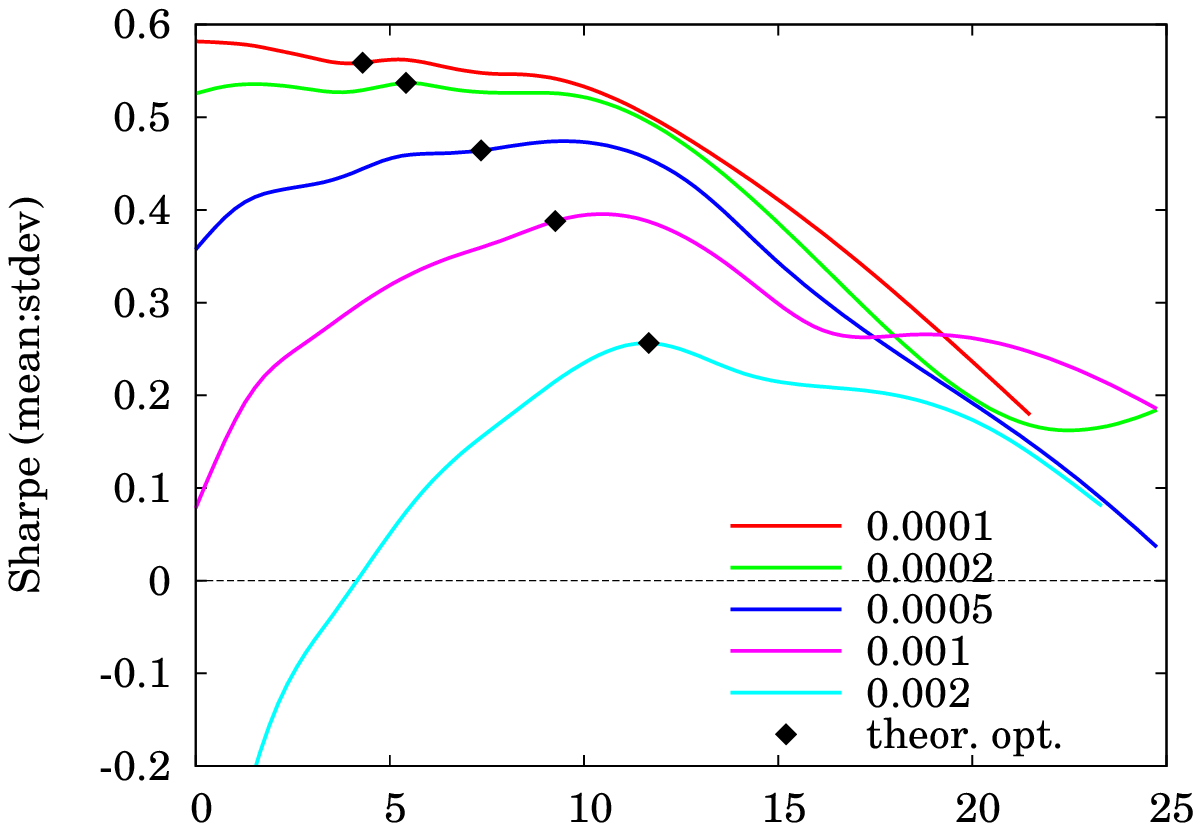}}
&
(d) \scalebox{0.55}{\includegraphics*{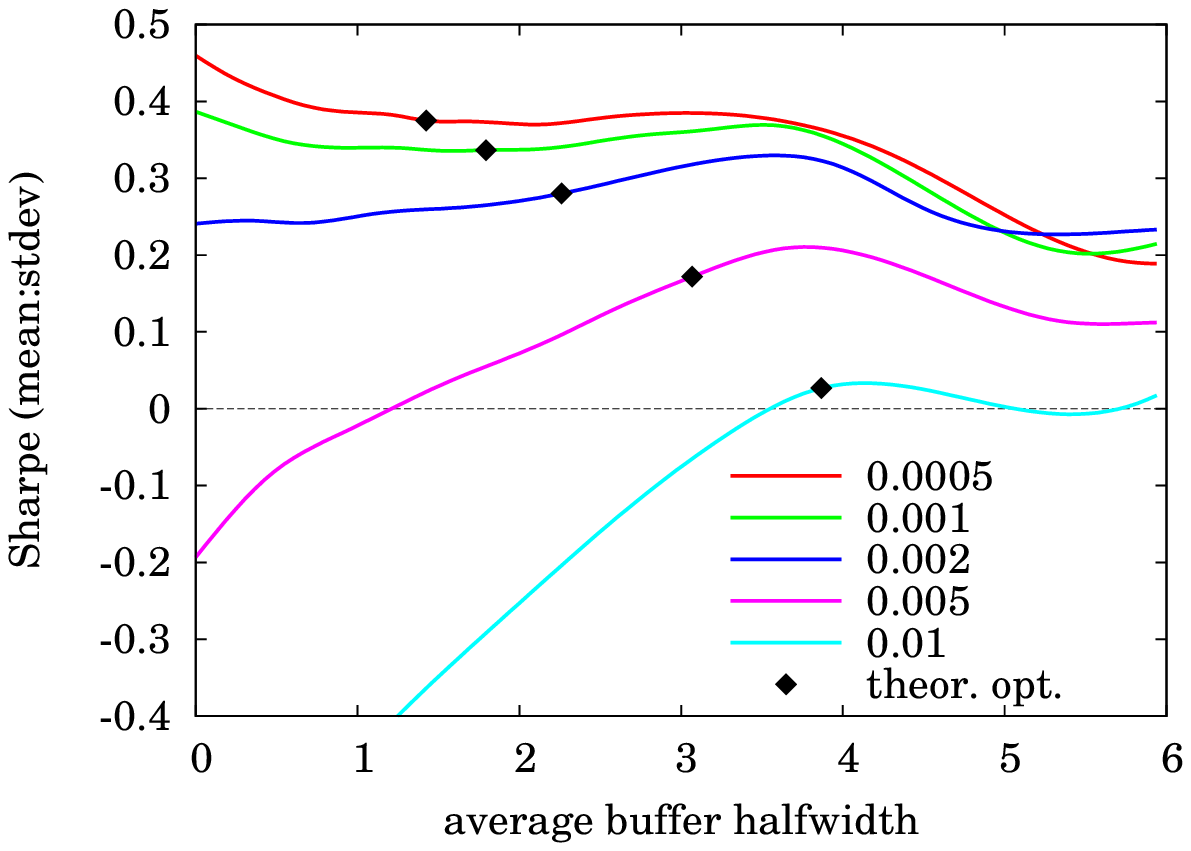}} \\
\\
(b) \scalebox{0.55}{\includegraphics*{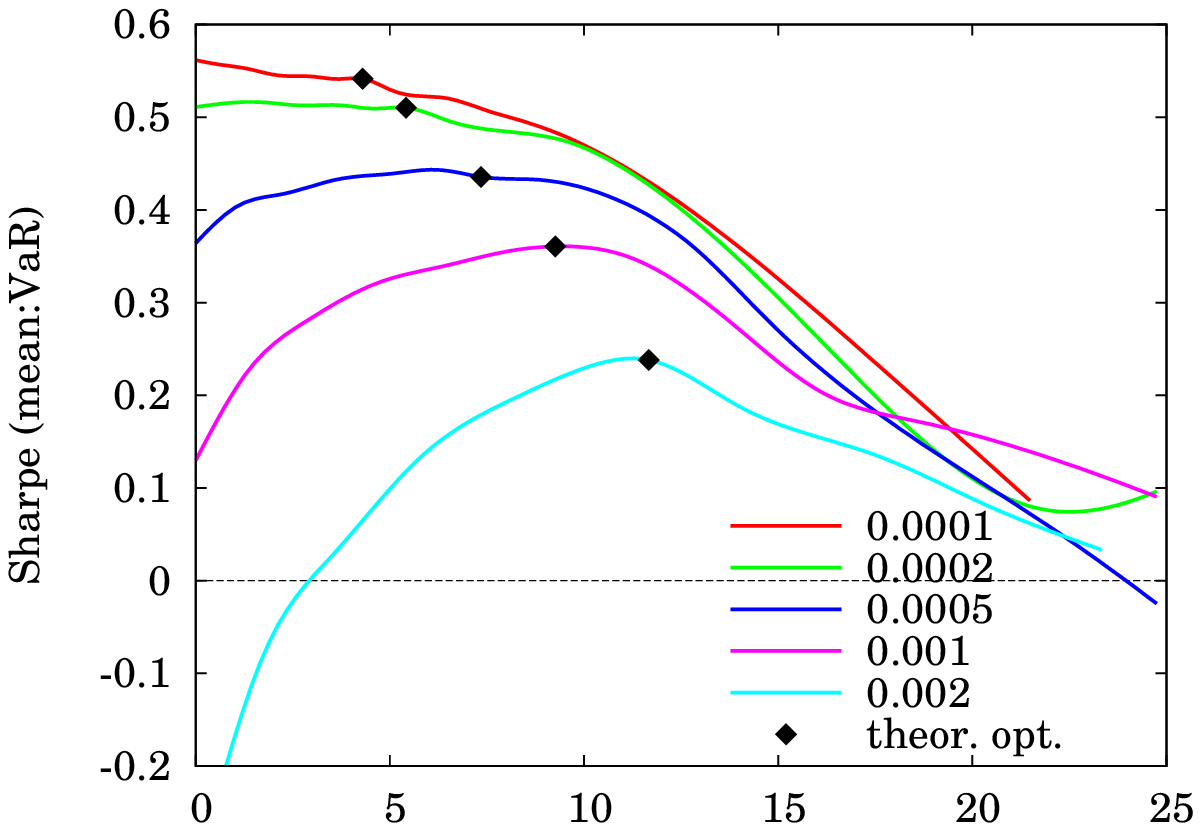}}
&
(e) \scalebox{0.55}{\includegraphics*{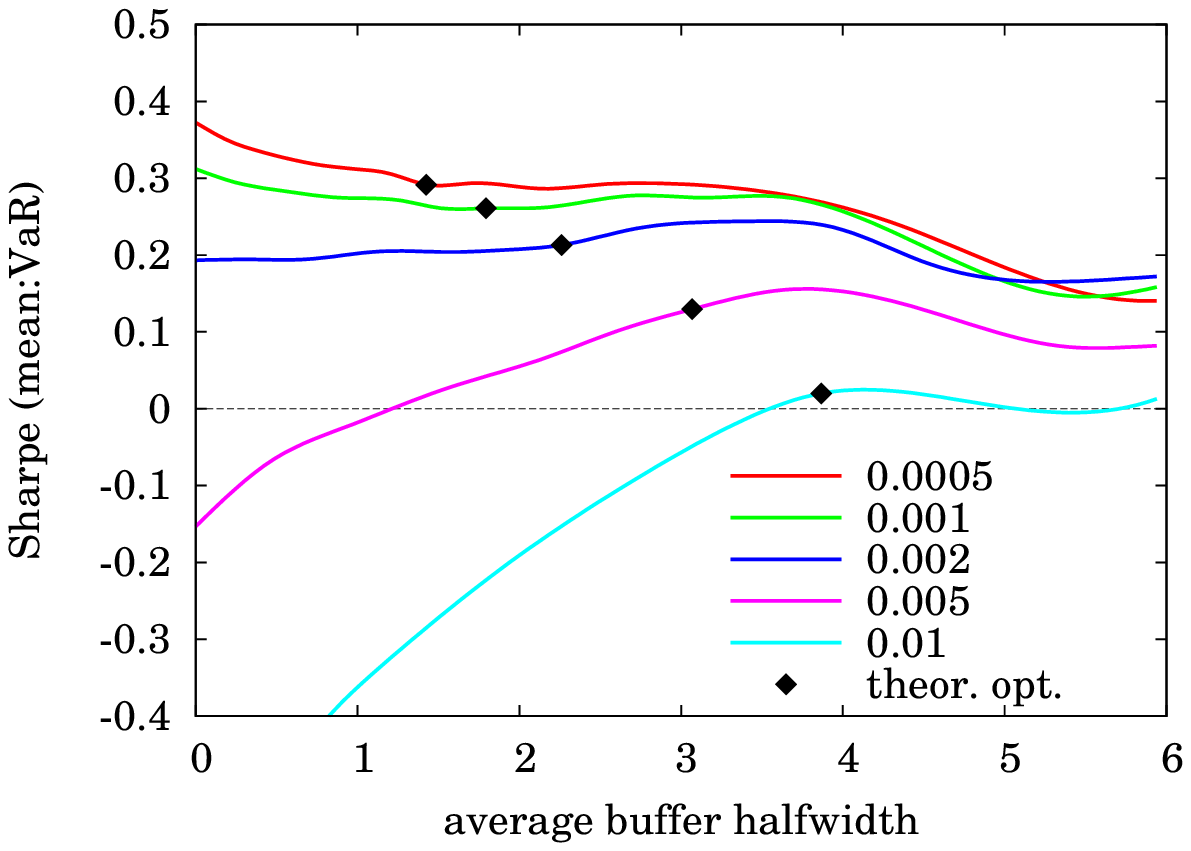}} \\
\\
(c) \scalebox{0.55}{\includegraphics*{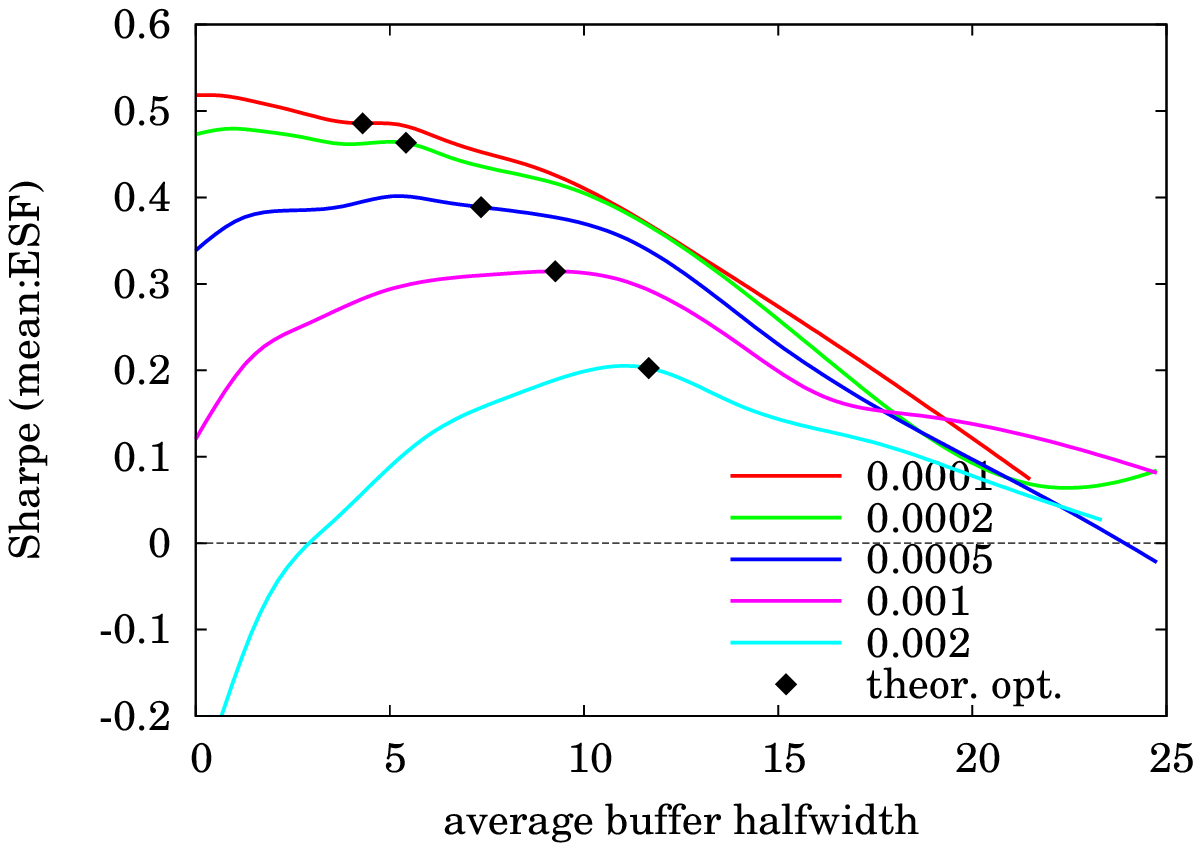}} 
&
(f) \scalebox{0.55}{\includegraphics*{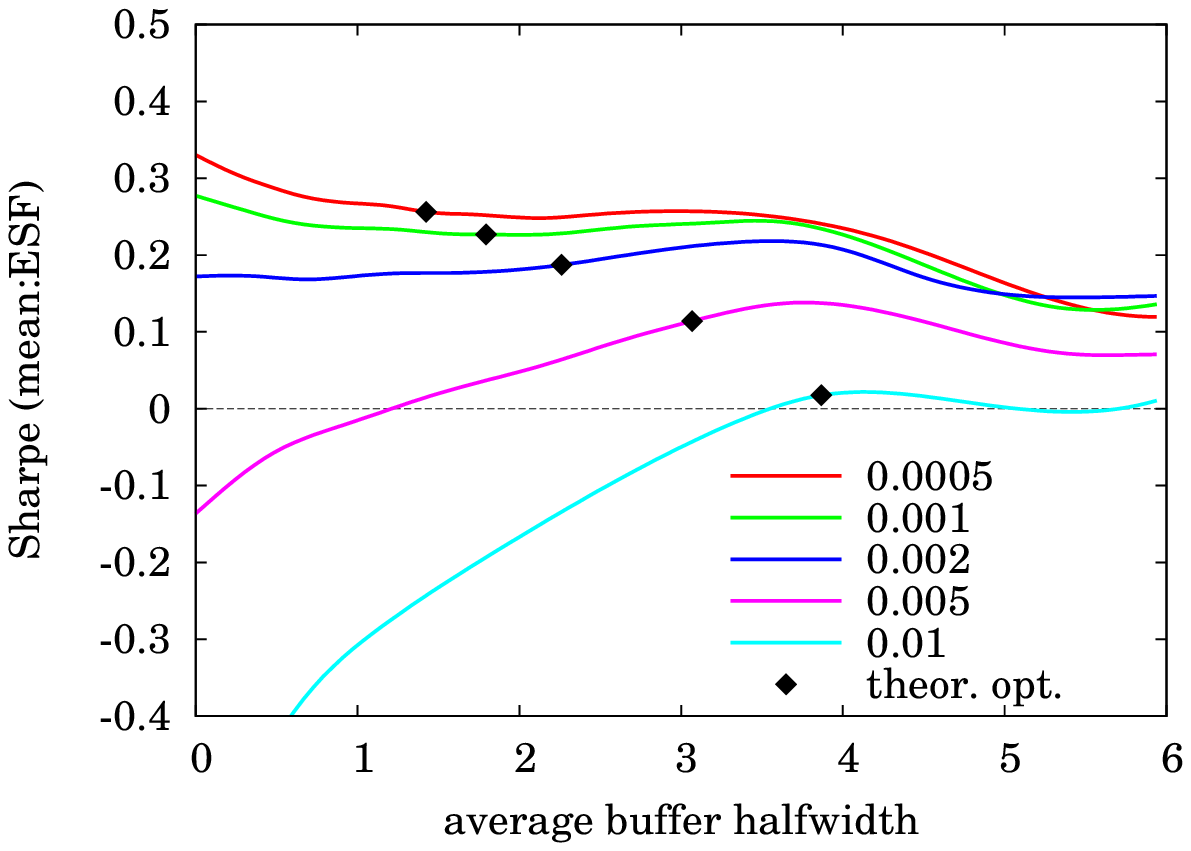}} 
\end{tabular}
\caption{\small Performance vs buffer size for momentum strategy, with different objective functions and markets. Cost multipliers ($\varepsilon$) are indicated on graph. Buffer size is as notional in \$M. Theoretical optimum marked in each case.
Objective functions: (a,d) Stdev, (b,e) VaR, (c,f) Shortfall. Markets: (a,b,c) US Treasury note TY1, (d,e,f) Rough rice (RR).
}
\label{fig:perf}
\end{figure}

The results are shown in Figure \ref{fig:perf}. 
The theoretical optimum is reasonably optimal in practice too, and again the effect of getting the buffer width wrong by a factor of 2 is substantial, except for very low costs. The same picture is seen for different risk measures.
For low transaction costs, the Sharpe ratio does not go negative in the limit of no buffering. This is because the simulations are being done in discrete time, an issue requiring further research. 
For rice, which has historically trended less well than bonds, notice that trading generates no value for high transaction costs, rendering the strategy ineffective.



\clearpage 


\section{Conclusions}

We have demonstrated a rule (\ref{eq:width}) for the optimal buffer, or NT, width to be applied to a diffusive factor model in the presence of proportional transaction costs and it seems to work well. For low costs\footnote{Strictly, this means lower transaction cost per unit volatility} it seems to slightly overestimate the optimal width in the `real' examples we showed, and we think this is due in part to the time discretisation in the simulation (the theory is continuous-time).

Clearly it is important to know whether a strategy can make money after costs, even if it is profitable in theory. Knowing how to correctly buffer a strategy is important when the transaction cost is high, as we have seen. If, despite optimising the model parameters and incorporating the buffer rule, the strategy's simulated expected return is still negative, then one knows to avoid it. One can also see even before simulating whether costs are infeasibly high: the buffer gets so big that the model exhibits too much hysteresis, getting stuck in the same position for perhaps months or years, and is effectively inoperable.

\subsection*{Acknowledgement}

The author thanks all of the following: Torsten Sch\"oneborn (Deutsche Bank), Chris Rogers (University of Cambridge), Jean-Philippe Bouchaud, J\'er\^ome de Lataillade and Rapha\"el B\'enichou (CFM, Paris), Mete Soner and Johannes Muhle-Karbe (ETH Z\"urich), Boris Gnedenko (Modern Investment Technologies Ltd), and an anonymous referee.
 
\bibliographystyle{plain}
\bibliography{../phd}

\end{document}